\documentstyle[epsf]{mn}
\twocolumn
\newcommand{\beq}{\begin{equation}}
\newcommand{\eeq}{\end{equation}}
\newcommand{\beqa}{\begin{eqnarray}}
\newcommand{\eeqa}{\end{eqnarray}}

\newcommand{\lexp}{\mathop{\langle}}
\newcommand{\rexp}{\mathop{\rangle}}

\def\d{\delta}

\def\del{\nabla}

\def\x{{\bf x}}
\def\q{{\bf q}}
\def\v{{\bf v}}
\def\la{\mathrel{\mathpalette\fun <}}
\def\ga{\mathrel{\mathpalette\fun >}}
\def\fun#1#2{\lower3.6pt\vbox{\baselineskip0pt\lineskip.9pt
\ialign{$\mathsurround=0pt#1\hfill##\hfil$\crcr#2\crcr\sim\crcr}}}

\title{PTHalos:  A fast method for generating mock galaxy distributions}

\author[R. Scoccimarro and R. K. Sheth]
{Rom\'{a}n Scoccimarro$^{1}$ and Ravi K. Sheth$^{2}$  \\ 
${}^1$Department of Physics, New York University, 4 Washington Place, 
New York, NY 10003 \\  
${}^2$NASA/FermiLab Astrophysics Center, P.O.Box 500, Batavia, IL 60510}

\pagerange{000,000}

\begin{document}
\maketitle

%
\begin{abstract}
%

Current models of galaxy formation applied to understanding the
large-scale structure of the universe have two parts.  The first is an
accurate solution of the equations of motion for the dark matter due
to gravitational clustering.  The second consists of making physically
reasonable approximations to the behavior of baryons inside dark
matter halos. The first uses large, computationally intensive,
$n$-body simulations.  We argue that because the second step is, at
least at present, uncertain, it is possible to obtain similar 
galaxy distributions without solving the first step exactly.

We describe an algorithm which is several orders of magnitude faster
than $n$-body simulations, but which is, nevertheless, rather
accurate.  The algorithm combines perturbation theory with virialized
halo models of the nonlinear density and velocity fields.  For two-
and three-point statistics the resulting fields are exact on large
scales, and rather accurate well into the nonlinear regime,
particularly for two-point statistics in real and redshift space. We
then show how to use this algorithm to generate mock galaxy
distributions from halo occupation numbers.  As a first application,
we show that it provides a good description of the clustering of
galaxies in the PSCz survey.

We also discuss applications to the estimation of non-Gaussian
contributions to error bars and covariance matrix of the power
spectrum, in real and redshift space, for galaxies and dark
matter. The results for the latter show good agreement with
simulations, supporting the use of our method to constrain
cosmological parameters from upcoming galaxy surveys.

\end{abstract}


%
%
\section{Introduction}
\label{intro}
%
%

Upcoming large galaxy surveys will provide a major advance in our
understanding of the large-scale structure of the universe, helping to
determine cosmological parameters, constrain models of galaxy 
formation and the properties of primordial
fluctuations that gave rise to galaxies. However, extracting
information about cosmological parameters from the galaxy distribution
requires an accurate modeling of non-linear effects such as
gravitational evolution, redshift distortions, and galaxy biasing
(i.e., the relationship between the galaxy and the underlying dark
matter distribution). 

The physics of galaxy formation is not yet fully understood.
Numerical simulations which include the effects of both gravitational
instability as well as gas hydrodynamics in cosmological volumes are
only now becoming available (e.g. Cen \& Ostriker 2000; Pearce et
al. 2001). But the role of feedback from star formation, and how to
incorporate it into simulations, remains uncertain.  As a result, a
complementary approach to generating realistic galaxy distributions
has been to use semianalytic models (see, e.g., Kauffmann et al. 1999;
Somerville \& Primack 1999; Cole et al. 2000).

These models build on the work of White \& Rees (1978) and White \&
Frenk (1991), in which galaxy formation is treated as a two-stage
process: dark matter haloes virialize, and gas cools and forms stars
within these virialized halos.  The first step is solved numerically:
$n$-body simulations of nonlinear gravitational clustering are used to
follow the formation of the dark matter halos.  The second step uses a
number of reasonable prescriptions for approximating the complicated
nonlinear physics of gas cooling in gravitational potential wells to
incorporate star and galaxy formation into the simulations which
otherwise only describe the effects of gravitational clustering.  The
predictions of the semi-analytic models, while similar to those from
smoothed particle simulations which solve the hydrodynamic equations
for the gas numerically, can differ by as much as fifty percent in the
two-point correlation function (e.g. Benson et al. 2001).  Therefore,
it is this second step which is the more uncertain one, since we do
not yet understand galaxy formation from first principles.  On the
other hand, it is the first step which is the most time consuming.

In this paper we present a method which generates realistic galaxy
distributions in a very small fraction of the time it takes for
methods that require $n$-body simulations of gravitational
clustering. By realistic distributions, we mean specifically that the
point distributions our method generates can reproduce the observed
galaxy clustering statistics; in particular, the distribution of
galaxy counts-in-cells, the power spectrum and the bispectrum. The
philosophy of our approach is that since the step which involves
generating galaxies from knowledge of the dark matter distribution is
necessarily uncertain, one does not need to start with a fully correct
dark matter distribution to generate galaxy correlations to the 
extent allowed by the present understanding of galaxy formation.
By suitably approximating the non-linear structures which form in the
gravitational clustering simulations, one can hope to still obtain a
reasonably accurate galaxy distribution by slightly altering the mapping from
dark matter to galaxies within the uncertainties.

To obtain an approximation to the fully nonlinear dark matter
distribution we first generate the large-scale dark matter
distribution using second-order Lagrangian perturbation theory (2LPT).
This reproduces the correct two and three-point statistics at large
scales, and approximates the four-point and higher-order statistics
very well (Moutarde et al. 1991; Buchert et al. 1994; Bouchet et
al. 1995; Scoccimarro 2000).  Note that use of 2LPT, rather than
linear theory or the Zel'dovich approximation, is essential to
incorporate accurately the large-scale departures from Gaussian
initial conditions.

The 2LPT correlations are incorrect on small scales, where
perturbation theory breaks down.  We build up more accurate
small-scale correlations by using the amplitude of the 2LPT density
field to determine the masses and positions of virialized halo
centers, and we then distribute particles around the halo centers with
realistic density profiles.  We use the 2LPT code described by
Scoccimarro (1998,2000) to set up the perturbation theory density and
velocity fields, and the merger history algorithm of Sheth \& Lemson
(1999b) to partition the 2LPT density field into haloes.  Finally, a
galaxy distribution can be generated by specifying how many galaxies
populate haloes of a given mass (e.g. Mo, Jing \& B\"orner 1997; Jing,
Mo \& B\"orner 1998).  In this respect, our {\tt PTHalos} algorithm
treats the difference between the clustering statistics of the dark
matter and galaxy distributions in much the same way that recent
analytic models (Peacock \& Smith 2000; Seljak 2000; Scoccimarro et
al. 2001) do.

In this paper, we do not attempt to match other galaxy properties
(e.g. the luminosity function) than clustering statistics.  Statistics
like the luminosity function can be estimated rather easily by simply
replacing the dark matter distribution of an $n$-body simulation with
the {\tt PTHalos} distribution, and then running the usual
semianalytic galaxy formation codes on top.  Our motivation was mainly
to develop a fast algorithm which can be used to generate realistic
non-Gaussian galaxy distributions, and thus provide a method to
quickly explore parameter space for constraining cosmological
parameters and galaxy formation models from galaxy surveys. In
addition, the speed of our code makes feasible to construct a large
number of mock galaxy catalogs in a reasonable CPU time from which
reliable estimation of errors and covariance matrices can be
derived. Application along these lines to imaging data in the SDSS
survey will be considered elsewhere (Connolly et al. 2001; Dodelson et
al. 2001; Scranton et al. 2001; Szalay et al. 2001; Tegmark et
al. 2001).

This paper is organized as follows. In Section~\ref{2lpt} we review
2LPT.  In Section~\ref{halos} we describe how we parametrize haloes
and in Section~\ref{pthalos} how we place them in the 2LPT density
field. In Section~\ref{nbody}, we compare the clustering statistics of
the dark matter distribution of {\tt PTHalos} to $n$-body simulations.  In
Section~\ref{gal} we describe how we map from dark matter to galaxies.
We show that our method provides a sensible match to to the galaxies
in the PSCz survey. We conclude in Section~\ref{concl}.

%
%
\section{Second-Order Lagrangian PT}
\label{2lpt}
%
%

In Lagrangian PT, the dynamics is described by the displacement field
${\bf \Psi}(\q)$ which maps the initial particle positions $\q$ into
the final Eulerian particle positions $\x$,

\beq
\x = \q + {\bf \Psi}(\q).
\eeq

\noindent The equation of motion for particle trajectories $\x(\tau)$
is

\beq
\frac{d^2 \x}{d \tau^2} + {\cal H}(\tau) \ \frac{d \x}{d \tau}= -
\del \Phi, 
\eeq

\noindent where $\Phi$ denotes the gravitational potential, and $\del$
the gradient operator in Eulerian coordinates $\x$. Taking the
divergence of this equation we obtain

\beq
J(\q,\tau)\ \del \cdot \Big[ \frac{d^2 \x}{d \tau^2} + {\cal
H}(\tau) \ \frac{d \x}{d \tau} \Big] = \frac{3}{2} \Omega_m {\cal H}^2
(J-1),
\label{leom}
\eeq

\noindent where we have used Poisson equation together with the fact
that $1+\d(\x) =J^{-1}$, and the Jacobian $J(\q,\tau)$ is the
determinant

\beq
J(\q,\tau)  \equiv  {\rm Det}\Big( \d_{ij}+ \Psi_{i,j} \Big),
\label{jacobian}
\eeq

\noindent where $\Psi_{i,j} \equiv \partial\Psi_i /\partial
\q_j$. Equation~(\ref{leom}) can be fully rewritten in terms of
Lagrangian coordinates by using that $\del_i = ( \d_{ij}+
\Psi_{i,j})^{-1} \del_{q_j}$, where $\del_q \equiv \partial /\partial
\q$ denotes the gradient operator in Lagrangian coordinates. The
resulting non-linear equation for ${\bf \Psi}(\q)$ is then solved
perturbatively, expanding about its linear solution, the Zel'dovich
(1970) approximation

\beq
\del_q \cdot {\bf \Psi}^{(1)}= -D_1(\tau) \ \d(\q).
\label{Psi1}
\eeq 

\noindent Here $\d(\q)$ denotes the (Gaussian) density field imposed
by the initial conditions and $D_1(\tau)$ is the linear growth
factor. The solution to second order describes the correction to the
ZA displacement due to gravitational tidal effects and reads

\beq
\del_q \cdot {\bf \Psi}^{(2)}= \frac{1}{2} D_2(\tau) \sum_{i \neq j}
(\Psi_{i,i}^{(1)} \Psi_{j,j}^{(1)} - \Psi_{i,j}^{(1)} \Psi_{j,i}^{(1)}),
\label{Psi2}
\eeq

\noindent (e.g., Bouchet et al. 1995) where $D_2(\tau)$ denotes the
second-order growth factor, which for flat models with non-zero
cosmological constant $\Lambda$ we have for $0.01 \leq \Omega_m \leq
1$

\beq
D_2(\tau) \approx -\frac{3}{7} D_1^2(\tau) \ \Omega_m^{-1/143} \approx
-\frac{3}{7} D_1^2(\tau),  
\eeq

\noindent to better than 0.6\% and 2.6\%, respectively (Bouchet et
al. 1995).  Since Lagrangian solutions up to second-order are
curl-free, it is convenient to define Lagrangian potentials
$\phi^{(1)}$ and $\phi^{(2)}$ so that in 2LPT

\beq
\x(\q) = \q -D_1\ \del_q \phi^{(1)} + D_2\ \del_q \phi^{(2)},
\label{dis2}
\eeq

\noindent and the velocity field then reads ($t$ denotes cosmic time)

\beq
{\bf v} \equiv \frac{d \x}{d t} = -D_1\ f_1\ H\ \del_q \phi^{(1)}
+ D_2\ f_2\ H\ \del_q \phi^{(2)},
\label{vel2}
\eeq 

\noindent where $H$ is the Hubble constant, and the logarithmic
derivatives of the growth factors $f_i \equiv (d\ln D_i)/(d \ln a)$
can be approximated for flat models with non-zero cosmological
constant $\Lambda$ and $0.01 \leq \Omega_m \leq 1$

\beq
f_1 \approx \Omega_m^{5/9}, \ \ \ \ \ f_2 \approx 2 \ \Omega_m^{6/11},
\eeq 

\noindent to better than 10\% and 12\%, respectively (Bouchet et
al. 1995). The accuracy of these two fits improves significantly for
$\Omega_m \geq 0.1$, in the range relevant according to present
observations.  The time-independent potentials in Eqs.~(\ref{dis2})
and~(\ref{vel2}) obey the following Poisson equations (Buchert et
al. 1994)
\label{poisson2lpt}
\beqa
\del_q^2 \phi^{(1)}(\q) &=&  \d(\q), 
\label{phi1} \\ 
\del_q^2 \phi^{(2)}(\q)&=&  \sum_{i>j}
[\phi_{,ii}^{(1)}(\q)\ \phi_{,jj}^{(1)}(\q) - (\phi_{,ij}^{(1)}(\q))^2].
\label{phi2}
\eeqa 

Thus, 2LPT positions and velocities can all be determined from the
initial fluctuation field.  In practice, we generate 2LPT positions
and velocities using the algorithm described in detail in Appendix~D
of Scoccimarro (1998), where the Poisson equations are solved by
standard fast Fourier transform methods. As mentioned before, 2LPT
recovers the exact two and three-point statistics at large scales, and
approximates very well higher-order ones (see e.g. Fig.~15 in Bouchet
et al. 1995, and Table~2 in Scoccimarro 2000). It is possible to
improve on 2LPT by going to third-order in the displacement field
(3LPT), however it becomes more costly due to the need of solving
three additional Poisson equations (Buchert 1994). 3LPT reproduces
exactly up to four-point statistics, and improves the behavior in the
underdense regions (where 2LPT tends to overestimate the density, see
Bouchet et al. 1995).

\begin{figure*}
\epsfxsize=9truecm
\epsfysize=9truecm
\centerline{\epsffile{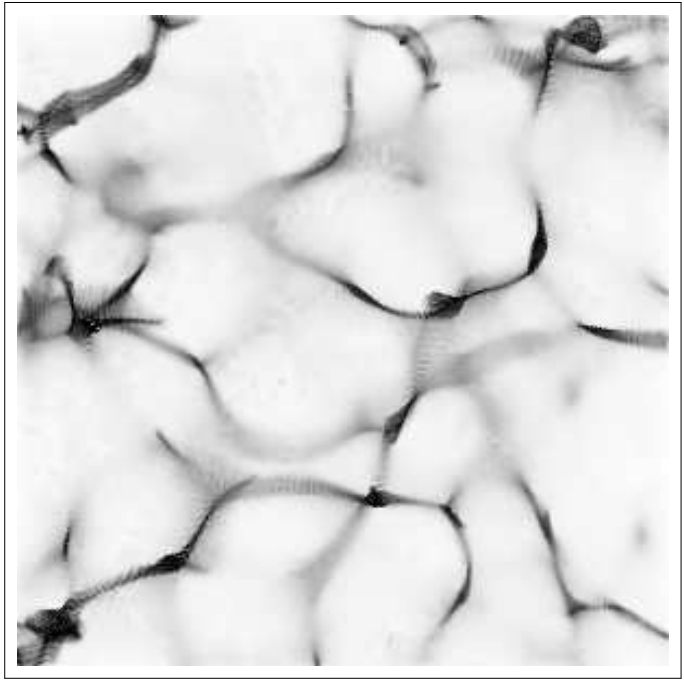}}
\caption{A slice of 2LPT, 150 Mpc/h a side and 6 Mpc/h thick.}
\label{fig2lpt}
\end{figure*}
\begin{figure*}
\epsfxsize=9truecm
\epsfysize=9truecm
\centerline{\epsffile{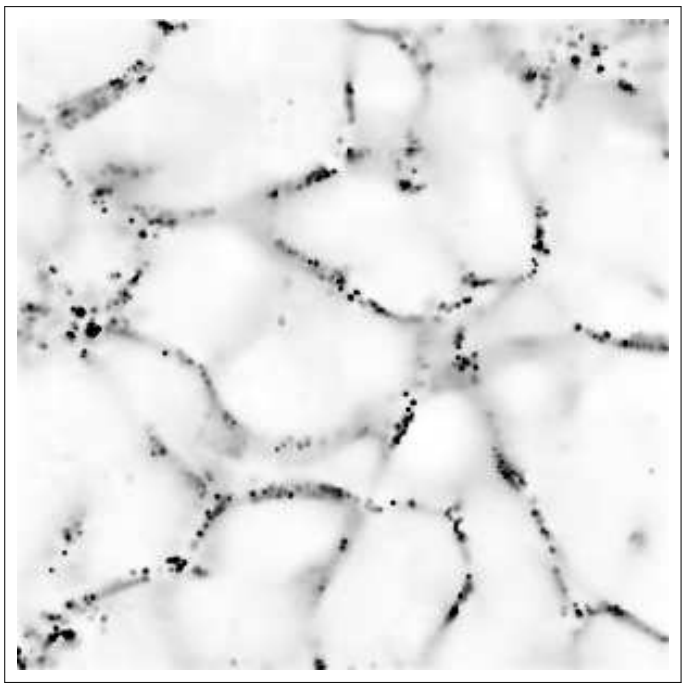}}
\caption{The same slice as previous figure for {\tt PTHalos}.}
\label{figPTHalos}
\end{figure*}

%
%
\section{From 2LPT to virialized haloes}\label{halos}
%
%

The end result of 2LPT is a list of particle positions and velocities.  
These positions and velocities are not quite the same as those the 
particles would have had in a full $n$-body simulation which started 
from the same initial conditions.  This section describes how to use 
our knowledge of fully nonlinear density and velocity fields to 
increase the agreement with simulations.

The primary difference between the 2LPT density field and that from 
a full $n$-body simulation of nonlinear gravitational clustering is 
that the 2LPT density field has no virialized objects.  In a full 
simulation, however, most of the mass in a simulation box is 
partitioned into virialized haloes (e.g. Tormen 1998).  For our 
purposes here, all virialized objects have well defined edges, and 
the edge, called the virial radius, is defined so that all virialized 
haloes have the same spherically averaged density, whatever their 
mass:  $m\propto r_{\rm vir}^3$.  Therefore, in what follows, we will 
make the simplifying assumption that all virialized haloes are 
spherical.  (In fact, haloes have a range of shapes; allowing for a 
distribution of shapes is a detail which does not change the logic 
of our method, so we will comment on it in more detail later.)

We will approximate the effects of fully nonlinear clustering in two 
steps.  First, we need a prescription for dividing up the 2LPT density 
field into a collection of virialized haloes.  Once this has been done, 
we must decide how to distribute the mass associated with each halo 
around the halo centre-of-mass.  This second step is more straightforward, 
so we will describe it first.  

\subsection{Virialized halo profiles}
As noted above, we will assume that all dark matter haloes are spherically 
symmetric.  High resolution $n$-body simulations 
(Navarro, Frenk \& White 1997) show that the spherically 
averaged density run around the centre of a virialized halo containing 
mass $m$ within the virial radius $r_{\rm vir}$ is well fit by 
\begin{equation}
{\rho(r)\over\bar\rho} = 
  {\Delta_{\rm vir}(z)\over 3\Omega(z)}{c^3f(c)\over x(1+x)^2},
        \qquad{\rm where}\ x \equiv c(m)\,{r\over r_{\rm vir}}, 
\label{rhonfw}
\end{equation}
where $\bar\rho(z)$ is the average density of the background universe 
at $z$, $\bar\rho(z)\,\Delta_{\rm vir}(z)/\Omega(z)$ is the average 
density within the virial radius ($\Delta_{\rm vir}\approx 178$ for 
all $z$ in an Einstein de-Sitter universe; it is $\approx 102$ at $z=0$ 
for the $\Lambda$CDM model we present results for in this paper), and 
$f(c) = [\ln(1+c)-c/(1+c)]^{-1}$.
The density run is a broken power-law, with a shallower inner slope 
and a steeper outer slope.  The exact shape of these two slopes is the 
subject of some debate.  Although we will use the NFW form in what 
follows, it is trivial to modify our algorithm to generate profiles of 
the form given, e.g., by Hernquist (1990) or by Moore et al. (1999).  
The parameter $c$ is often called the central concentration of the 
halo.  As NFW noted, more massive haloes are less centrally 
concentrated.  We will use the parametrization of this trend provided 
by Bullock et al. (2001):  
\beq
c(m) \approx {9\over (1+z)} \Big( \frac{m}{m_{*0}} \Big)^{-0.13} ,
\label{conc}
\eeq where $m_{*0}$ is the standard non-linear mass scale (defined in
the next subsection). Numerical simulations show that not all haloes
of mass $m$ have the same density profile, there is considerable
scatter. Fortunately, this scatter is well fit by using the same NFW
shape for all haloes, but letting the concentration parameter have a
Lognormal distribution with dispersion $\Delta(\log c) \approx 0.2$
(Jing 2000; Bullock et al. 2001).  We include this scatter in
{\tt PTHalos}.

Equation~(\ref{rhonfw}) says that if the centre of mass of an $m$-halo 
is at position $\x$, then there will be $N_{\rm dm}\propto m$ particles 
distributed around $\x$ according to equation~(\ref{rhonfw}).  
This is easily done.  For example, if the halo profile were an 
isothermal sphere ($\rho \propto r^{-2}$), then particle positions 
around the halo centre $(r,\theta,\phi)$ could be got by drawing random 
numbers distributed uniformly between zero and $r_{\rm vir}$ for the 
radial distance $r$ from the centre, uniform numbers between plus 
and minus one for $\cos\theta$, and uniform numbers between zero and 
two pi for $\phi$.  Generating an NFW profile is only slightly more 
complicated.  

Because the density run depends on halo mass $m$, the trick is to 
identify those positions in the 2LPT density field which should be 
identified with the centres-of-masses of $m$-haloes.  
The next subsection describes how to do this.  

\subsection{Halo masses and positions}
Imagine comparing the 2LPT density and velocity fields with those 
from an $n$-body simulation which started from the same initial 
fluctuation field.  One might imagine that the 2LPT density and 
velocity fields contain information about where bound objects in 
the $n$-body simulation formed.  For example, perhaps the densest 
2LPT regions are those regions which, in the $n$-body simulation, 
collapsed to form bound haloes.  If so, then one might imagine running 
a friends-of-friends group finder on the 2LPT density field to identify 
these overdense regions.  One could then use the number of particles 
associated with the friends-of-friends group as an estimate of the 
mass of the virialized halo which formed in the $n$-body simulation, 
and the position of the centre-of-mass of the group could be used as 
an estimate of the position of the corresponding halo.  
While such a procedure is possible in principle, running a group-finder 
can be quite time-consuming.  This is the primary reason why we have 
adopted the approach we describe below.  

At any given time, which we will label by redshift $z$, virialized 
haloes have a wide range of masses.  Let $n(m|z)$ denote the comoving 
number density of haloes of mass $m$ at $z$.  The shape of this universal 
mass function distribution is well approximated by
\begin{equation}
{m^2\,n(m|z)\over \bar\rho} = \nu f(\nu)\,
                               {{\rm d}{\rm ln}\nu\over {\rm d}{\rm ln}m},
\label{m2nm}
\end{equation}
where $\bar\rho$ denotes the comoving density of the background, and 
\begin{equation}
\nu f(\nu) = 2A\Big(1+(a\nu^2)^{-p}\Big) 
     \left(a\nu^2\over 2\pi\right)^{1/2}\,\exp\left(-{a\nu^2\over 2}\right).
\label{vfvgif}
\end{equation}
Here $\nu \equiv \d_{\rm sc}(z)/\sigma(m,z)$ where $\d_{\rm sc}(z)$ is
the collapse threshold given by the spherical collapse model (e.g. it
is 1.68 in an Einstein-deSitter universe), $\sigma^2(m,z)$ is the
linear theory variance in the density field when smoothed on the
comoving scale $R = (3m/4\pi\bar\rho)^{1/3}$ at $z$, and
$A=0.5,0.322$, $p=0,0.3$ and $a=1,0.707$ for the mass functions given
by Press \& Schechter (1974) and Sheth \& Tormen (1999), respectively.
(Sheth, Mo \& Tormen 2001 show that these two cases may be related to
models in which objects form from spherical or ellipsoidal collapses,
respectively.  Jenkins et al. (2001) compare these mass functions with
$n$-body simulations.)  The number of haloes falls exponentially at
large masses.  Let $m_*$ denote the mass at which this cut-off sets
in.  Then $m_*$ is defined by requiring
$\sigma(m_*,z)\equiv\delta_{\rm sc}(z)$.  The quantity $m_{*0}$ which
should be used in equation~(\ref{conc}) is got from setting
$\sigma(m_{*0},0)\equiv\delta_{\rm sc}(0)$.

The distribution of haloes in dense regions is different from that 
in underdense regions:  $n(m,\delta|z) \ne (1+\delta)\,n(m|z)$, 
where $\delta$ denotes the overdensity of the region.  
For example, the ratio of massive to less massive haloes is larger 
in dense regions than in underdense ones.  A simple model for this 
dependence is 
\begin{equation}
n(m,\delta|z) \approx \Bigl[1+b(m|z)\,\delta\Bigr]\,n(m|z),
\end{equation}
where $\delta \equiv M/\bar\rho V - 1$, and 
\begin{equation}
b(m|z) = 1 + {a\nu^2-1\over\delta_{\rm sc}(z)} + 
         {2p/\delta_{\rm sc}(z)\over 1 + (a\nu^2)^p} 
\end{equation}
(Mo \& White 1996; Sheth \& Tormen 1999).  
The actual detailed dependence of $n(m,\delta|z)$ on $\delta$ can 
be computed following Lemson \& Kauffmann (1999), Sheth \& Lemson (1999a) 
and Sheth \& Tormen (2001), so we will not repeat the analysis here.  

If we place a grid on the 2LPT density field, then some of the cells 
will be denser than others.  We must find an algorithm which ensures 
that the distribution of halo masses in the different density cells 
follows the correct $n(m,\delta|z)$ relation.  

We can do this as follows.  The actual nonlinear density in $V$ at $z$
is given by $M/V\equiv \bar\rho(1+\delta)$.  Let $\delta_0$ denote the
value for the overdensity one would have predicted for such a region,
had one used linear theory to make the prediction.  Typically, we
expect that if $\delta > 0$, then $\delta_0 < \delta$, and viceversa
for underdense regions.  A simple fitting formula to the relation
between the nonlinear and the linear overdensities, $\delta$ and
$\delta_0$, got from assuming that objects form from a spherical
collapse in an Einstein-deSitter universe, has been provided by Mo \&
White (1996).  We have checked that the following simple modification
to their formula is accurate for all cosmologies of interest:
\begin{eqnarray}
\delta_0\!\!\! &=&\!\!\! \left[1.68647 - {1.35\over (1+\delta)^{2/3}} - 
{1.12431\over (1+\delta)^{1/2}} + {0.78785\over (1+\delta)^{0.58661}}\right]
\nonumber \\
&&\qquad\qquad\qquad \times\qquad {\delta_{\rm sc}(z)\over 1.68647}.
\label{d0mow}
\end{eqnarray}
Now, what we are trying to do is to partition the mass $M$ in $V$ at 
$z$ up into subregions, each of nonlinear density 
 $\Delta_{\rm vir}(z)\gg 1$, 
and hence each with predicted linear density $\delta_{\rm sc}(z)$.  

This is extremely similar to what one does when one studies the 
merger histories of objects.  Given the mass $M$ and the linear 
theory density $\delta_0$ associated with that mass, one studies 
the distribution of subclumps $m_j$ of $M$ at some earlier time when 
the critical density for collapse was $\delta_{\rm sc}(z)$.  
A number of merger history tree codes which do this are available.  
The algorithm described by Sheth \& Lemson (1999b) is simple and 
efficient, so we will use it to partition the 2LPT mass $M$ up into 
virialized haloes.  The only difference is that, whereas they were 
trying to generate the Press-Schechter mass function, we would like 
to generate the distribution which fits the simulations better.  
This can be done by making the following simple change to their 
algorithm.  

\begin{figure}
\centering 
\epsfxsize=\hsize\epsffile{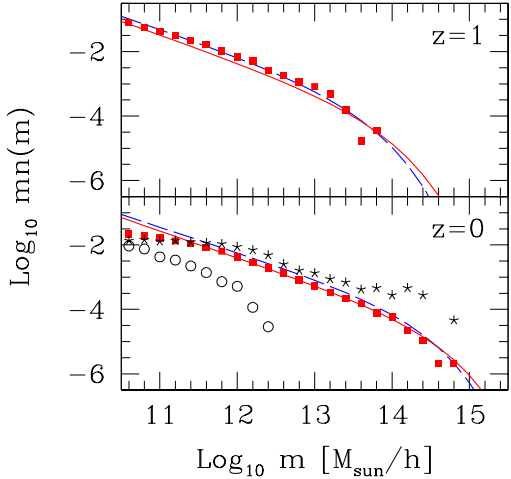}
\caption{The halo mass function at $z=1$ (top) and $z=0$ (bottom).  In
each panel, squares show the output from {\tt PTHalos}$\!\!$, solid line
shows the fitting function which describes the $n$-body simulations,
and dashed line shows the Press--Schechter mass function.  Stars and
circles in the bottom panel show the mass function in the densest and
the least dense three percent of the box at $z=0$. These results
correspond to flat $\Lambda$CDM realizations ($\sigma_8=0.9$,
$\Omega_m=0.3$ at $z=0$) in a 100 Mpc/h box with mass resolution
$m_{\rm min}=10^{10.5} M_{\sun}$/h and $R_{\rm grid}=12.5$ Mpc/h.  }
\label{mnmgif}
\end{figure}

Sheth \& Lemson's algorithm loops over a range of `time' steps, 
chooses a series of Gaussian random numbers $g_i$ at each time step, 
and then, for each $i$, makes use of $g_i^{-2}$.  Their algorithm is 
fast because there are efficient ways to generate Gaussian variates.  
Therefore, if we modify the algorithm, we would like to do it in such 
a way that it still requires only Gaussian variates.  For $\Lambda$CDM 
we have found that requiring the number of `time' steps to equal ten, 
and then using  $\ a\,[3 + 0.33/(1 + |g_i|^{1.5})]/|g_i|$ instead of 
$g_i^{-2}$, where $a=0.707$ is the same parameter as in 
equation~(\ref{vfvgif}), is all that is required to generate a mass 
function which is more like the $n$-body simulations (see Fig.~\ref{mnmgif}).  

Notice that because $(1+\delta) \le \Delta_{\rm vir}$, 
equation~(\ref{d0mow}) guarantees that $\delta_0\le \delta_{\rm sc}(z)$.  
Therefore, in essence, by using a merger tree code to perform the 
partition, we are making use of the following fact: 
a dense cell can be thought of as a region which will, at some point 
in the near future, become a virialized object.  Therefore, when we 
view a dense cell at the present time, it is as though we are 
viewing a virialized object at a small `lookback time' from the 
time it virialized.  At small lookback times, most of the mass $M$ of 
an object is likely to be partitioned up into just a few pieces which 
are each a substantial fraction of $M$, because there has not been enough 
time for $M$ to have been split up into many smaller pieces.  In 
hierarchical clustering models, the opposite is true at large lookback 
times.  This provides a simple reason why the ratio of massive to less 
massive haloes is larger in dense regions than in underdense ones.  
In particular, this shows that when we partition the mass of a cell up 
as though we were partitioning the mass of a virialized halo up into 
subhaloes at high redshift, the only decision to be made is which 
redshift, which `lookback time', to choose when running the merger tree 
algorithm.  We have argued above that this choice depends on $\delta$:  
the exact transformation is given by equation~(\ref{d0mow}).  

To summarize, given the mass $M$ and the 2LPT overdensity $\delta$ in 
a cell $V$, a merger tree algorithm is used to split $M$ up into virialized 
haloes $m_j$.  The next step is to decide where within $V$ to place 
the halo centres.  We do this by assuming that the most massive haloes 
within $V$ occupy the densest subregion within $V$.  Once this has been 
done, all that remains is to distribute particles around each halo centre 
as described in the previous subsection.  

\subsection{Nonlinear velocities}
The final step is to account for the differences between the 2LPT and
$n$-body velocity fields.  In the $n$-body simulations, it is a good
approximation to assume that the motion of a particle can be written
as the sum of two terms: 
\beq \v = \v_{\rm vir} + \v_{\rm halo} , 
\eeq
where the first term represents the virial motion of the particle
within its parent halo, and the second term is the bulk motion of the
halo as a whole (Sheth \& Diaferio 2001).  Furthermore, the virial
motions within a halo are well approximated by velocities which are
independent Gaussians in each of the three cartesian components, with
rms values which depend on halo mass: $\sigma_{\rm vir}^2=\langle
v_{\rm vir}^2\rangle \propto Gm/r_{\rm vir} \propto m^{2/3}$. In
particular, we use (Bryan \& Norman 1998; Sheth \& Diaferio 2001)

\begin{equation}
\sigma_{\rm vir} = 476 f_{\rm vir} (\Delta_{nl} E(z)^2)^{1/6} 
\Big( \frac{m}{10^{15} M_{\sun}/h} \Big)^{1/3} \ {\rm km/s},
\label{sigmavir}
\end{equation}

\noindent where $f_{\rm vir}=0.9$ and $\Delta_{nl}=18\pi^2+60 x-32
x^2$ with $x=\Omega(z)-1$, $\Omega(z)=\Omega_0 (1+z)^3/E(z)^2$,
$E(z)^2=\Omega_0 (1+z)^3+\Omega_\Lambda$ for a flat model with
cosmological constant.

We assume that virial motions are uncorrelated with the direction or
amplitude of $\v_{\rm halo}$.  Therefore, if we substitute the 2LPT
velocity vector at the position of the halo centre-of-mass for
$\v_{\rm halo}$, and then add an uncorrelated vector of virial motions
to each particle, and we do this for each of the particles in the
halo, then the resulting velocity field should be quite similar to
that in the $n$-body simulations. This is essentially a numerical
implementation of the analytical model developed in Sheth et al
(2001).

This is the final step in converting the 2LPT density and velocity 
fields to fields which resemble the $n$-body simulations more closely.  
The next section describes the algorithm which implements all of this.

%
%
\section{PTHalos}
\label{pthalos}
%
%

\subsection{The Algorithm}

Given a realization of the large-scale density field, {\tt PTHalos} assigns
halo centers to appropriate 2LPT particles, and then generates NFW
density profiles around them. This is done as follows.

\begin{itemize}

\item The desired mass resolution $m_{\rm min}$, the size of the
simulation volume, and the cosmological model set the abundance of
halos of mass larger than $m_{\rm min}$, $N_{\rm halos}$. 
Since halo centers are going to be identified as 2LPT particles, 
this sets the number of 2LPT particles to be used, 
$N_{\rm 2LPT} \approx N_{\rm halos}$, and thus their mass $m_{\rm 2LPT}$. 

\item The 2LPT box is divided into cubic cells of size $R_{\rm grid}$,
and the mass in 2LPT particles $M_i$ in each cell $i$ is obtained. The
choice of $R_{\rm grid}$ is dictated by the competing requirements
that it be small enough so that the mass distribution is not
rearranged at large scales (to preserve the correct correlations
imposed by 2LPT), and large enough so that, within the cell, there can
be halos sufficiently massive. Unless otherwise noted, we use $R_{\rm
grid}= 15$ Mpc/$h$. This corresponds roughly to a spherical cell of 8 
Mpc/h radius which is the natural non-linear scale. 

\item Then, a merger history code is run on each cell $i$, which 
gives a partition of the cell mass $M_i$ into halos of smaller 
mass $m_j$ ($M_i=m_1+m_2+...$) down to the mass resolution.  

\item The 2LPT particles in a cell and the list of halos resulting
from the merger history tree are matched so that most massive halos
are centered about 2LPT particles in the densest regions. An exclusion
volume of 2Mpc/$h$ (1Mpc/$h$ at $z=1$) radius is imposed about 2LPT
centers in densities larger than 5.6 (set by the turnaround density in
the spherical collapse model).  Halos of mass $m_j$ are constructed
about their 2LPT centers by sampling the NFW profile with the
appropriate number of particles according to the mass resolution.

\item Velocities are assigned using $\v=\v_{\rm 2LPT}+\v_{\rm vir}$,
where $\v_{\rm 2LPT}$ is the velocity of the 2LPT particle
representing the halo center of mass, and $\v_{\rm vir}$ is a virial
velocity drawn at random from a Maxwellian distribution with
one-dimensional dispersion $\sigma_{\rm vir}(m_j)$ given by
Eq.~\ref{sigmavir}.  We require that the center of mass velocity of
each halo equals that of the associated $\v_{\rm 2LPT}$; this is
always possible because $N_{\rm 2LPT} \ga N_{\rm halos}$. This
maintains the correct 2LPT correlations between density and velocity
fields at large scales, which is important if we wish to have the
correct large-scale redshift-space statistics.

\item A galaxy distribution can be obtained by specifying how many
galaxies, $N_{\rm gal}(m_j)$ on average populate dark matter halos of
mass $m_j$.  Given this function, we use a binomial distribution (as
explained in detail in Scoccimarro et al.  2001) to sample the NFW
profile with the resulting number of galaxies.  Benson et al.  (2000)
describe one possible alternative to the binomial.  The binomial model
we use has two free functions, the first and second moments of the
number of galaxies per halo of mass $m$.  The first moment is
parametrized by a broken power-law with a low-mass cutoff, as
described in Section~\ref{gal}; the second moment is related to the
first taking into account that low-mass halos have a sub-Poisson
dispersion (Kauffmann et al.  1999, Benson et al.  2000).

\end{itemize}

Fig.~\ref{fig2lpt} shows a slice of 150~Mpc/$h$ a side and 6~Mpc/$h$
thick from a 2LPT distribution corresponding to $\Lambda$CDM with
$\Omega_m=0.3$, $\Omega_\Lambda=0.7$, $\sigma_8=0.90$.
Figure~\ref{figPTHalos} shows the same slice in {\tt PTHalos}, after 2LPT
high-density regions have been replaced by virialized halos.  

\subsection{Memory Requirements and Speed}

The {\tt PTHalos} code requires a memory of 50 bytes per particle in the
simulation box. Most importantly, a $300^3$ particle realization takes
only about 5 minutes to generate on a single-CPU workstation, about
$2-3$ orders of magnitude faster than $n$-body simulations. For
particle numbers $N_{\rm par} \ga 300^3$ approximately equal time is
spent generating the 2LPT density field, imposing the spatial
exclusion of massive halos, generating profiles and velocities, and
writing the output files to disk. From tests we have performed up to
$N_{\rm par}=400^3$, the run time scales roughly as $N_{\rm par} \ln
(N_{\rm par})$.

\begin{figure}
\epsfxsize=9truecm
\epsfysize=9truecm
\centerline{\epsffile{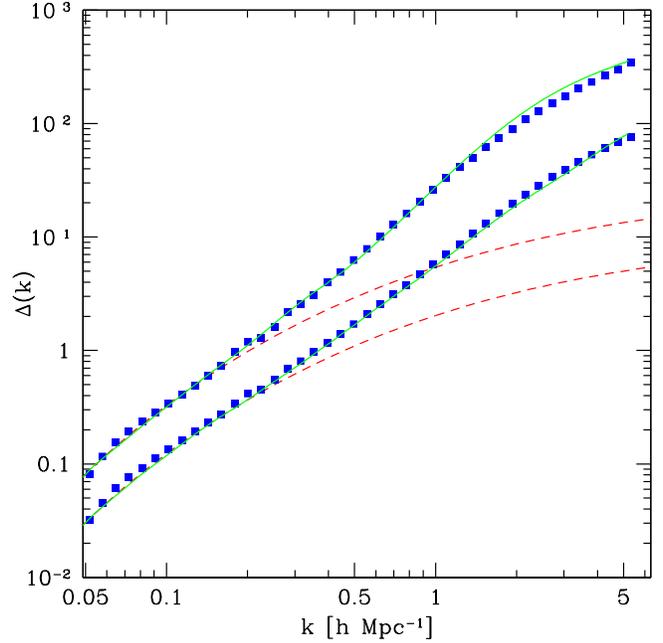}}
\caption{The power spectrum as a function of scale for $\Lambda$CDM at
$z=0$ (top) and $z=1$ (bottom). Symbols denote measurements in N-body
simulations, solid lines correspond to {\tt PTHalos} (averaged over 30
realizations), and dashed lines show linear perturbation theory.}
\label{figpk}
\end{figure}

\begin{figure}
\epsfxsize=9truecm
\epsfysize=9truecm
\centerline{\epsffile{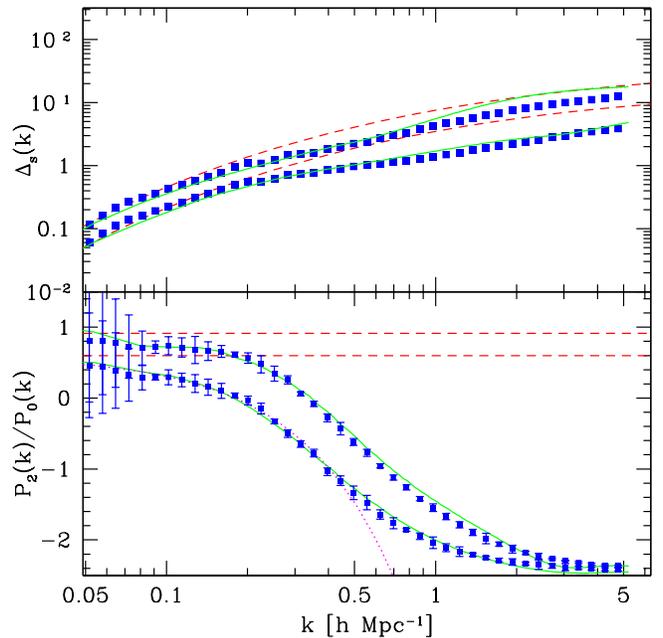}}
\caption{The power spectrum in redshift-space at $z=0$ and $z=1$. The
top panel shows the power spectrum monopole, the bottom panel
corresponds to the quadrupole to monopole ratio. Line styles are as in
Fig~\protect{\ref{figpk}}. The dotted line in the bottom panel shows
the fitting formula by Hatton \& Cole (1999).}
\label{figpkz}
\end{figure}

\begin{figure}
\epsfxsize=9truecm
\epsfysize=9truecm
\centerline{\epsffile{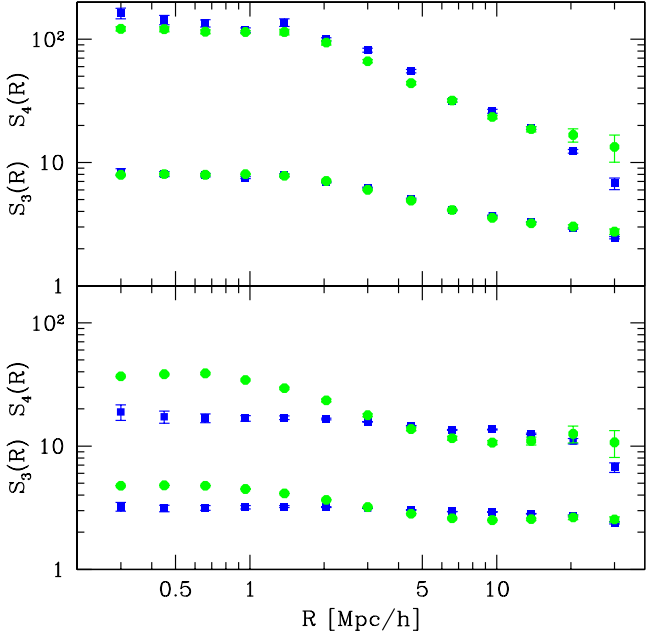}}
\caption{The skewness ($S_3$) and kurtosis ($S_4$) as a function of
smoothing scale $R$ in real (top) and redshift (bottom) space. Circles
denote measurements in ten {\tt PTHalos} realizations, and squares
correspond to the average over four $n$-body simulations. }
\label{figSp}
\end{figure}

%
%
\section{PTHalos vs. Numerical Simulations}
 \label{nbody}
%
%

We now turn to a quantitative comparison of clustering statistics for
the dark matter obtained from {\tt PTHalos} to $n$-body simulations. 
Application of {\tt PTHalos} to the distribution of PSCz
galaxies is discussed in Section~\ref{gal}.

\subsection{Mass Function}

A first test of our algorithm is to check that we obtain the correct
mass function of dark matter halos.  Fig.~\ref{mnmgif} shows the
number density of haloes as a function of halo mass, for a simulation
box of side 100 Mpc/h, mass resolution $m_{\rm min}=10^{10.5}
M_{\sun}/h$ and $R_{\rm grid}=12.5$ Mpc/h.  The symbols show the
universal mass function one gets after using the merger tree algorithm
to partition the 2LPT mass into haloes, and the solid curves show the
fitting formula which describes the universal mass function in
$n$-body simulations (equations~\ref{m2nm} and~\ref{vfvgif}).  The
symbols match the curves quite well, suggesting that our decision to
use a merger tree algorithm was quite successful.  The stars and
circles in the $z=0$ panel show the mass function in the densest and
the least dense three percent of the box.  This shows that the densest
cells contain the most massive haloes, as expected.

\subsection{Power Spectrum in Real and Redshift Space}

Figure~\ref{figpk} shows the power spectrum for the $\Lambda$CDM model
at $z=0$ ($\sigma_8=0.90$, $\Omega_m=0.3$, $\Omega_\Lambda=0.7$) and
$z=1$ ($\sigma_8=0.55$).  The symbols in this and next figure denote
the measurements in $n$-body simulations taken from Scoccimarro,
Couchman \& Frieman (1999a).  The simulation was run by the Virgo
Consortium; it corresponds to a single realization with $256^3$
particles in a 240 Mpc/h box.  The dashed lines show the linear power
spectrum, and the solid lines correspond to the measured power
spectrum averaged over 30 {\tt PTHalos} realizations. 
Figure~\ref{figpkz} shows similar measurements for the power spectrum
in redshift-space.  The top panel corresponds to the monopole at $z=0$
(top) and $z=1$ (bottom) and the bottom panel shows the quadrupole to
monopole ratio at $z=0$ (bottom) and $z=1$ (top).  Error bars for the
simulations correspond to the dispersion among four different lines of
sight for the redshift-space mapping, done in the plane-parallel
approximation, as assumed throughout this paper.

We see that the agreement between {\tt PTHalos} and numerical
simulations is very good, especially considering the simplicity of our
algorithm.  Similar results hold for the SCDM model.  In redshift
space, the power spectrum monopole is somewhat overestimated (for
$z=0$) at small scales, suggesting perhaps our model for virial
velocities is oversimplified (consistent with the analytic results
presented in Sheth et al.  2001).  Another possibility is inaccurate
treatment of halo spatial exclusion effects; in redshift space
halo-halo correlations are important down to smaller scales than in
real space, because velocity dispersion suppresses halo profiles.  We
shall come back to this point in the next subsection when we study
higher-order moments.

The excellent agreement shown for the quadrupole to monopole ratio
even at scales below zero-crossing is very encouraging, since until
now there was no alternative to numerical simulations that could
provide an accurate treatment of redshift distortions. The problem, as
emphasized by Hatton \& Cole (1998) is that phenomenological models,
where linear perturbation theory predictions (Kaiser 1987) are
convolved with a kernel which describes the effects of velocity
dispersion (Peacock \& Dodds 1994, Park et al. 1994), tend to
underestimate non-linear effects at large scales. Similarly, recent
models of redshift distortions based on the halo model (White 2001,
Seljak 2001) which assume linear perturbation theory for halo-halo
correlations do not accurately reproduce the quadrupole moment,
although they provide a good description of the power spectrum
monopole (White 2001).

In our case the large-scale ($k \la 0.2$ h/Mpc) behavior of the
monopole and quadrupole is dictated by 2LPT and has little to do with
velocity dispersion (unlike in phenomenological models), and is
reproduced very accurately. In fact, the bottom panel in
Fig.~\ref{figpkz} shows as dotted line the fitting formula found by
Hatton \& Cole (1999) empirically from an ensemble of $n$-body
simulations. This relation, valid at scales larger than that of the
quadrupole zero-crossing, fits our results extremely well. A detailed
explanation of how perturbation theory can accurately describe the
large-scale behavior of power spectrum multipoles will be presented
elsewhere.  Essentially, the large-scale pairwise velocity along the
line of sight is strong enough to suppress the power spectrum monopole
and drive the quadrupole to zero.  The demonstration of this effect
requires an accurate treatment of the non-linear nature of the
redshift-space mapping (Scoccimarro et al.  1999a), which in our
numerical treatment is trivially implemented since 2LPT treats the
dynamics perturbatively but the mapping to redshift space is done
exactly.  Note that the deviations from linear perturbation theory are
{\em not} negligible, even at scales larger than the non-linear scale
$k \approx 0.2$ h/Mpc, in particular for the quadrupole.

\subsection{Higher-Order Moments in Real and Redshift Space}

Figure~\ref{figSp} shows a comparison of {\tt PTHalos} against
$n$-body simulations for higher-order moments of the density field in
real (top) and redshift (bottom) space. The figure shows the skewness
$S_3(R)$ and kurtosis $S_4(R)$ as a function of smoothing scale
$R$. Squares denote the average over 4 realizations of $\Lambda$CDM
with $128^3$ particles in a box of $L_{\rm box}=300$ Mpc/h a side,
obtained by running the Hydra code (Couchman et al. 1995). The circles
denote the average of 10 realizations of {\tt PTHalos} with $300^3$
particles in the same volume.

The agreement in real space is very good, particularly for the
skewness.  The runs in this figure were made using $R_{\rm grid}=20$
Mpc/h, we found that for $R_{\rm grid}=15$ Mpc/h, the skewness and
kurtosis were underestimated at small scales $R<1$ Mpc/h by $10\%$ and
$70\%$, respectively. Changing $R_{\rm grid}$ does not affect the
power spectrum, but does affect increasingly more the higher-order
moments which are most sensitive to the presence of high-mass
halos. The choice of $R_{\rm grid}=15$ Mpc/h is thus a bit small for
$L_{\rm box}=300$ Mpc/h, which leads to an underestimation of the
number of very massive halos (which occasionally share more than one
cell 15 Mpc/h a side).

In redshift space (bottom panel in Fig.~\ref{figSp}) the situation is
somewhat different, the small-scale skewness and kurtosis are actually
higher in {\tt PTHalos} than in $n$-body simulations. Increasing
artificially the value of the velocity dispersion of halos,
Eq.~(\ref{sigmavir}), does not fix the discrepancy (since the skewness
and kurtosis are ratios of moments), suggesting perhaps that the
problem is due to our approximate treatment of halo exclusion. More
work is needed to fully understand the origin of this discrepancy.

\subsection{Power Spectrum Covariance Matrix}
\label{Pkcova}

As mentioned in the introduction, one of the motivations behind {\tt
PTHalos} is to provide a efficient tool for calculating accurate error
bars and covariance matrices for clustering statistics including the
effects of non-linear evolution, redshift distortions and galaxy
biasing, that can be used to constrain cosmological parameters, galaxy
formation models and the statistics of primordial fluctuations from
analysis of clustering in upcoming galaxy surveys.

Here we consider what impact these effects have on the error bars and
covariance matrix of the power spectrum.  As it is well known,
non-linear effects lead to increased error bars in individual band
power estimates and introduce correlations between them that are
absent in the Gaussian case (Meiksin \& White 1999; Scoccimarro,
Zaldarriaga \& Hui 1999b; Hamilton 2000).  In this Section we compare
the predictions of {\tt PTHalos} for the covariance matrix of the dark
matter power spectrum against the numerical simulation results in
Scoccimarro et al.  (1999b), and present results that include redshift
distortions and galaxy biasing as well.

All the covariance matrices are calculated using 300 realizations of
{\tt PTHalos}. The SCDM ($\Omega_m=1$, $\sigma_8=0.61$) realizations
contain $200^3$ particles in a box of side $L_{\rm box}=100$ Mpc/h
(identical volume and band power binning as the simulations we compare
to, with bin width $\delta k=2\pi/100$ h/Mpc). The $\Lambda$CDM
($\Omega_m=0.3$, $\Omega_\Lambda=0.7$, $\sigma_8=0.9$) realizations
contain $300^3$ particles in a box of side $L_{\rm box}=300$
Mpc/h. Finally, we construct realizations of galaxy distributions as
described above with halo occupation numbers given by Eq.~(\ref{Ngal}),
derived in the next section from comparison to the PSCz survey. These
have about $21.4\times 10^6$ galaxies in a box of side $L_{\rm
box}=300$ Mpc/h (we do not include the survey selection function, our
only purpose here is to explore the effects of galaxy weighing on the
covariance matrix). In these cases, the width of bins in k-space is
$\delta k=0.05$ h/Mpc

\begin{figure}
\epsfxsize=9truecm
\epsfysize=9truecm
\centerline{\epsffile{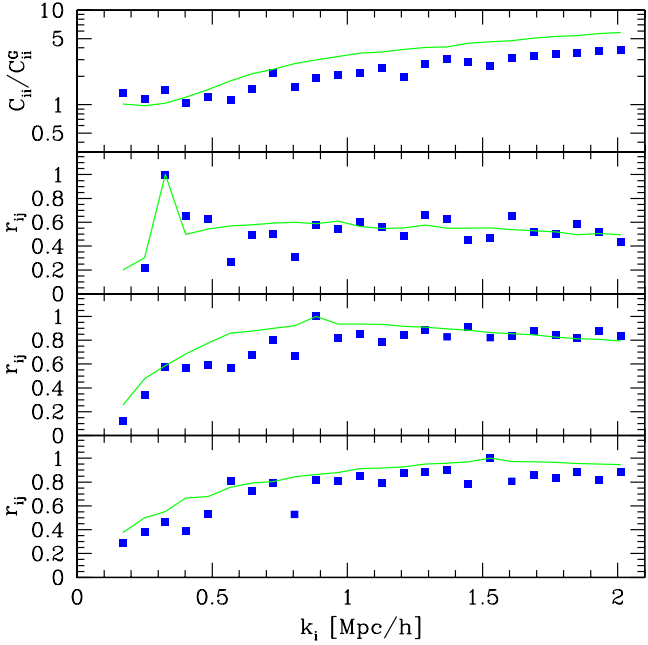}}
\caption{The Power Spectrum Covariance Matrix in SCDM $n$-body
simulations (symbols) and {\tt PTHalos} (solid lines). The top panel shows
the ratio of the power spectrum errors to those under the assumption
of Gaussianity. The remaining three panels show the cross-correlation
coefficient $r_{ij}$ between band powers centered at
$k_j=0.32,0.88,1.52$ h/Mpc as a function $k_i$ (Gaussianity
corresponds to $r_{ij}=\delta_{ij}$).}
\label{figPkcova}
\end{figure}

Figure~\ref{figPkcova} shows the results from {\tt PTHalos} (solid
lines) compared to the measurements obtained in Scoccimarro et al. 
(1999b) for the SCDM model.  The top panel shows the ratio of the power
spectrum errors (diagonal elements of the covariance matrix, $C_{ii}$)
to those under the assumption of Gaussianity, $C_{ii}^G$.  {\tt
PTHalos} seems to overestimate the errors at small scales by perhaps
as much as $50\%$, although we regard this as a preliminary result
since our $n$-body results contain only 20 realizations.  Indeed, from
our {\tt PTHalos} Monte Carlo pool we see that error bars estimated
from 20 realizations typically have a scatter of about $40\%$ compared
to the results from 300 realizations.  The remaining three panels in
Fig.~\ref{figPkcova} show the cross-correlation coefficient, $r_{ij}
\equiv C_{ij}/\sqrt{C_{ii}C_{jj}}$, between band powers centered at
$k_j=0.32,0.88,1.52$ h/Mpc as a function $k_i$.  Here the agreement
seems much better, although it seems {\tt PTHalos} slightly
overestimates cross-correlations.

\begin{figure}
\epsfxsize=9truecm
\epsfysize=9truecm
\centerline{\epsffile{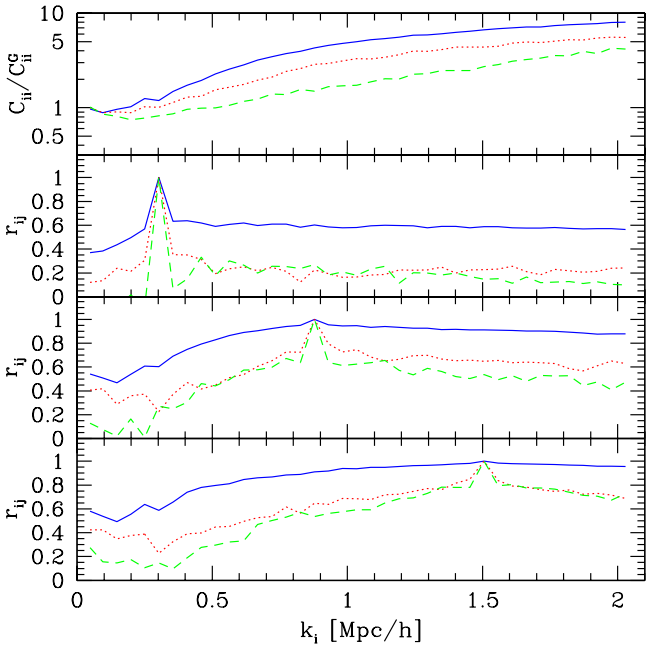}}
\caption{Same as Fig.~\protect{\ref{figPkcova}}, but for $\Lambda$CDM
{\tt PTHalos} realizations. Solid lines denote the real space dark matter
power spectrum covariance matrix, whereas dotted lines correspond to
redshift space. Dashed lines show redshift-space measurements of
galaxies given by Eq.~(\protect\ref{Ngal}).}
\label{figPkcova2}
\end{figure}

Figure~\ref{figPkcova2} shows results in the same format as
Fig.~\ref{figPkcova} but for $\Lambda$CDM {\tt PTHalos} realizations
for the dark matter power spectrum covariance matrix in real space
(solid lines), redshift space (dotted lines), and for galaxies [with
halo occupation numbers given by Eq.~(\ref{Ngal})] in redshift space
(dashed lines).  The effects of redshift distortions is to suppress the
non-Gaussian contribution to the errors and the cross-correlations
between band powers (Meiksin \& White 1999; Scoccimarro et al.  1999b),
due to the fact that velocity dispersion suppresses the higher-order
moments, see Fig.~\ref{figSp}.  For galaxies, there is the additional
effect of weighing the contribution of dark matter halos by the galaxy
occupation number.  At the scales shown here, this weighing suppresses
non-Gaussianity and thus errors and cross-correlations.  We find
however that for $k\ga 3$ h/Mpc (not shown), where the galaxy power
spectrum is larger than the dark matter power spectrum, the situation
reverses and galaxies have larger errors and cross-correlations than
dark matter (as can be guessed e.g. from the asymptotic behavior in
the top panel of Fig.~\ref{figPkcova2}).

Although galaxies in redshift space are generally less affected by
non-Gaussianity, the effect is strong enough to be very important for
constraining cosmological parameters from galaxy redshift surveys. 
Methods developed to decorrelate band powers (Hamilton 2000) were
shown to work very well for the mass power spectrum when
non-Gaussianities are modelled by the hierarchical model.  It would be
interesting to test these methods with {\tt PTHalos} galaxy
realizations to see how well they perform.

Another interesting application of our code would be to weak
gravitational lensing, where studies have been made of error bars and
covariance matrices for the power spectrum and moments of the
convergence field (White \& Hu 1999, Cooray \& Hu 2000, Van Waerbeke
et al 2001).

%
%
\section{Galaxy Distributions: Application to the PSCz survey}
\label{gal}
%
%

As discussed above, one the motivations behind {\tt PTHalos} is to
apply it to galaxy distributions, where its accuracy for galaxy
clustering is expected to be comparable to standard methods which are
computationally much more costly.

As a first example, we apply {\tt PTHalos} to clustering in the PSCz survey
(Saunders et al. 2000) assuming an underlying $\Lambda$CDM
($\Omega_m=0.3$, $\Omega_\Lambda=0.7$) cosmology. We use the following
measurements of clustering statistics: the real-space power spectrum
(Hamilton \& Tegmark 2001) inverted from redshift-space, the skewness
and kurtosis in redshift space as a function of scale (Szapudi et
al. 2000) and the redshift-space bispectrum (Feldman et al. 2001).

\begin{figure}
\epsfxsize=9truecm
\epsfysize=9truecm
\centerline{\epsffile{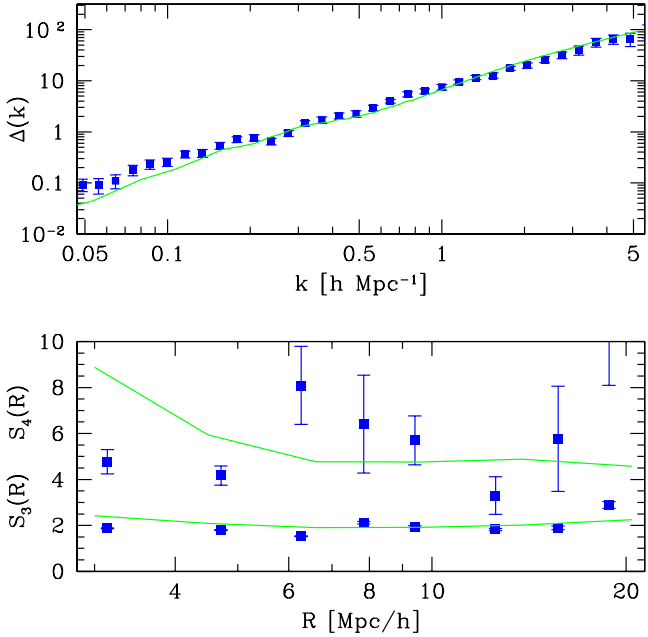}}
\caption{Top panel: real-space power spectrum inferred from the PSCz
survey (Hamilton \& Tegmark 2001) compared to the predictions for
galaxies obeying the relation in Eq.(\protect{\ref{Ngal}}). Bottom
panel: same for the skewness $S_3$ and kurtosis $S_4$ (Szapudi et al
2000).}
\label{figPkPSCz}
\end{figure}

In order to specify a galaxy distribution, we need to specify the
moments of the number of galaxies $N_{\rm gal}$ per halo of mass $m$. 
We use a binomial distribution, as described in Scoccimarro et al. 
(2001), which has only 2 free functions, the first and second moments
of $N_{\rm gal}(m)$.  We relate these two by using the semianalytic
model results of Kauffmann et al.  (1999), where the dispersion in
$N_{\rm gal}(m)$ is sub-Poisson for dark matter halos of mass
$m<10^{13}M_{\sun}$/h, and Poisson otherwise.  We then searched for a
first moment parametrized by a broken power-law that leads to a good
match with the real-space power spectrum of PSCz galaxies, and found

\beq
\lexp N_{\rm gal}(m) \rexp =0.7\ (m/m_0)^{\alpha},
\label{Ngal}
\eeq

\noindent where $\alpha=0$ for $8 \times 10^{10} M_{\sun}/h \leq m
\leq m_0$, $\alpha=0.7$ for $m > m_0$, $m_0=4\times 10^{11}
M_{\sun}/h$. By definition, $N_{\rm gal}=0$ for masses below the mass
resolution of the {\tt PTHalos} realization, $8 \times 10^{10}
M_{\sun}/h$. The resulting power spectrum is shown in the top panel in
Fig.~\ref{figPkPSCz} in solid lines, compared to the inferred power
spectrum from the PSCz survey, shown in symbols with error bars
(Hamilton \& Tegmark 2001). Jing, B\"orner and Suto (2001) found
recently that a similar relation, with $\alpha=0.75$, fits the
two-point correlation function of PSCz galaxies.

\begin{figure}
\epsfxsize=9truecm
\epsfysize=9truecm
\centerline{\epsffile{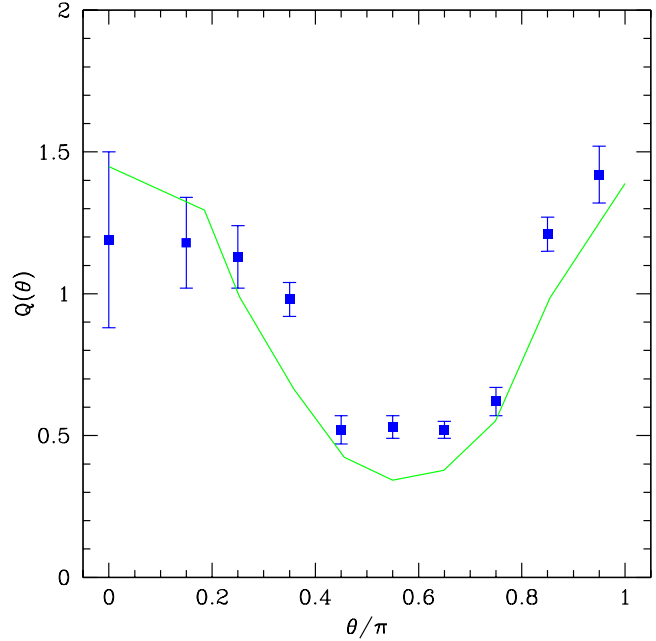}}
\caption{The redshift-space bispectrum in the weakly non-linear regime
($k<0.3$ h/Mpc) as a function of triangle shape for galaxies
populating halos as in Eq.(\protect{\ref{Ngal}}), compared to the PSCz
bispectrum as measured by Feldman et al.  (2001).  The plot
corresponds to wavectors in a ratio of approximately $k_{1}/k_{2}=0.5$
and angle between them equal to $\theta$.  Error bars here represent
the scatter in each bin from measurements of triangles of the same
shape but somewhat different overall scale, and should only be
considered a rough estimate of the uncertainties, see Feldman et al. 
(2001) for a full analysis.}
\label{figQPSCz}
\end{figure}

We then calculated the skewness, kurtosis, and bispectrum in
redshift-space (bottom panel in Fig.~\ref{figPkPSCz} and
Fig.~\ref{figQPSCz}). The results turned out to be in very good
agreement with the measurements in the PSCz survey (we only consider
scales larger than $R=3$ Mpc/h in Fig.~\ref{figPkPSCz} given that
{\tt PTHalos} is not yet accurate enough for higher-order moments in
redshift space, see Fig~\ref{figSp}).  This is very encouraging,
because we did not attempt to fit them by adjusting the relation in
equation~(\ref{Ngal}). A similar result has been found by Szapudi et 
al. (2000) using the semianalytic models of Benson et al. (2000).

The resulting bias parameters are also in good agreement with the full
likelihood analysis of the PSCz bispectrum (Feldman et al.  2001),
$1/b_1 \approx 1.2\pm 0.2$ and $b_2/b_1^2 \approx -0.4 \pm 0.2$. 
Indeed, from taking the ratio of the power spectrum of the {\tt
PTHalos} galaxies to the underlying dark matter spectrum at large
scales, $k<0.3$ h/Mpc, we find $1/b_1=1.28$.  Comparing the
large-scale skewness for the dark matter, $S_3=2.6$, and for the {\tt
PTHalos} galaxies, $S_3'=2$, and using $S_3'\approx
S_3/b_1+3b_2/b_1^2$ (Fry \& Gazta\~naga 1993), yields
$b_2/b_1^2=-0.44$.

We stress that the result in Eq.~(\ref{Ngal}) should not be considered
a full analysis of constraints on galaxy occupation numbers from
clustering of PSCz galaxies, since we have not explored the parameter
space for $N_{\rm gal}(m)$ in a detailed fashion and only considered
the constraint from the power spectrum measurements.  However, the
fact that this exercise led to very reasonable higher-order moments
in redshift space gives us confidence that {\tt PTHalos} can be
successfully used to constrain galaxy formation models by using
measurements of redshift-space clustering statistics at scales $R \ga
3$ Mpc/h.

%
%
\section{Conclusions}
\label{concl}
%
%

We have described {\tt PTHalos}, a fast algorithm which generates
point distributions which have similar clustering statistics to those
of the observed galaxy distribution.  We have shown that our approach,
although based on simple approximations, is able to reproduce the
results of numerical simulations very well and requires minimal
computational resources.  Its speed and flexibility makes it suitable
to explore parameter space in constraining cosmological parameters and
galaxy formation models that can be later studied in more detail if
necessary by perhaps more accurate (and computationally costly)
methods.

As we noted in the introduction, although we only used {\tt PTHalos}
to model galaxy clustering, if one uses the {\tt PTHalos} dark matter
distribution in place of an $n$-body simulation, then, by placing a
semianalytic galaxy formation package on top, one can model other
properties of galaxies, such as luminosities and colours, as well.  If
one is interested in how bulge-to-disk ratios, or the correlation
between galaxy luminosity and velocity dispersion, depend on local
density, then the models described by, e.g., Dalcanton, Spergel \&
Summers (1997), Mo, Mao \& White (1998), van den Bosch (2000), etc. 
can also be placed on top of the {\tt PTHalos} dark matter
distribution.  These models may differ substantially in their
formulation and in the specific physical processes which were
considered relevant.  The fact that {\tt PTHalos} is several orders of
magnitude faster than approaches which require the output of an
$n$-body simulation means that it offers an efficient way of testing
these different models.

There is an interesting history of making mock catalogs of the galaxy
distribution in which one does not start with an $n$-body simulation. 
Soneira \& Peebles (1978) describe an algorithm which generates point
distributions which have the same small-scale two-point statistics as
the observations.  However, their method does not allow them to
generate velocities and, because they were designed to work on small
scales only, they do not include large-scale correlations.  Motivated
by the fact that the observed galaxy distribution is approximately
Lognormal, Coles \& Jones (1991) suggested that it might be a
reasonable approximation to simply map initially Gaussian fluctuations
to non-Gaussian ones by setting $\exp(\delta_{\rm init}) =
1+\delta_{\rm final}$.  However, this mapping is only approximate, and
it is not obvious how one might assign velocities, or account for the
difference in clustering between the dark matter and galaxies.  In
addition, although it is possible to generate one-point higher-order
moments by a local transformation of a Gaussian field, this does not
generate the correct configuration dependence of multi-point
correlation functions, which in gravitational instability are the
result of the non-local character of Poisson's equation.

Sheth \& Saslaw (1994) described a halo-based algorithm which is
similar in spirit to the one presented here.  However, their
prescription had no large-scale correlations (their haloes were given
a Poisson spatial distribution), and they did not describe how to
incorporate velocities into their model, nor how to allow for
differences between the dark matter and the galaxy distributions.
Bond \& Myers (1996) described a halo-based algorithm which did
include spatial correlations and used the Zel'dovich approximation for
velocities.  However, their algorithm was rather computationally
intensive.  Moreover, although it produced approximately the right
number of massive halos, it was less accurate at the low mass end.

Our {\tt PTHalos} algorithm represents a real improvement because:  
\begin{itemize}
\item it has correct two- and three-point correlations on large-scales;
\item it has accurate higher-order moments on all scales; 
\item it includes realistic velocity correlations; and 
\item it allows one to model the differences between the distributions 
of dark matter and different galaxy types---nontrivial biasing effects 
are rather simple to incorporate.  
\end{itemize}

There are a number of inputs to {\tt PTHalos} which are easily modified.
For example, we distribute particles around halo centres so that they
follow an NFW profile; this is easily changed to one's favourite
profile---a tophat, an isothermal sphere, or one of the profiles
described by Hernquist (1990) or Moore et al. (1999).  Also, our code
assumes that all haloes are smooth, whereas haloes in simulations have
substructure.  Neglecting this fact is reasonable on all but the very
smallest scales because substructure accounts for only about fifteen
percent of the mass of a halo (e.g. Ghigna et al. 2000).  In any case,
Blasi \& Sheth (2000) provide simple fitting formulae to Ghigna et
al.'s simulations which make including substructure straightforward.

It is slightly more complicated to generate haloes with a range of
ellipticities.  There are, at present, no convenient parametrizations
of the distribution of halo shapes in $n$-body simulations (see
however Dubinski \& Carlberg 1992).  The ellipsoidal collapse model
which reproduces the correct halo mass function makes specific
predictions for this distribution (Sheth, Mo \& Tormen 2001); in
principle, it could be used to specify the distribution of shapes. The
simplification of spherical halos leads to discrepancies in the
bispectrum, at small scales the reduced bispectrum $Q$ looses the
dependence on triangle shape at scales larger than those of the
simulations (Scoccimarro et al. 2001). One the other hand, the
numerical results in Ma \& Fry (2000) suggest that except possibly for
the smallest scales, the low-order spatial statistics at least are
insensitive to whether or not the haloes are spherical.

Improvements which we hope to include in the near future include the
following.  One of the places in which {\tt PTHalos} is not as accurate as
we would like is in its assignment of virial velocities.  At present,
we assume that velocities within haloes are isotropic.  Cole \& Lacey
(1996) show that this is not quite correct; haloes in simulations tend
to have slightly more radial orbits near the edge of the halo.  This
is something we have included as a test but it turned out to be a
small effect in the statistics we studied. However, in view of the
deviations found for redshift-space higher-order moments, more work is
needed to make sure our velocity statistics are accurate enough at
small scales.  In this connection, most importantly perhaps is that
our present scheme which accounts for the fact that haloes may not
overlap is somewhat adhoc.  An improved treatment of exclusion, or a
better algorithm for using the 2LPT density field to assign positions
to the haloes output from the merger tree partition would be very
useful.  One promising option is to modify the approach described by
Bond \& Myers (1996), so that it can be applied to the 2LPT rather
than the linear fluctuation field, to assign positions to the most
massive halos.  One could then use the merger tree algorithm to assign
the mass which remains to less massive haloes---one of the advantages
of Sheth \& Lemson's (1999b) merger tree algorithm is that it is
easily adapted to construct partitions when it is known that some of
the mass has already been assigned to more massive haloes.

To model deep galaxy redshift surveys, the angular correlations of
galaxies, weak lensing observations, quasar absorption lines and the
Ly-$\alpha$ forest, or the thermal and kinematic Sunyaev-Zel'dovich
effects, rather than providing particle positions and velocities at a
fixed epoch, it is more useful to have a simulation output particle
positions and velocities along the observer's light-cone.  In {\tt
PTHalos} this is particularly straightforward, and is the subject of
ongoing work.

Finally, another issue that is important for generating mock galaxy
surveys is to adapt the simulation to the required survey geometry and
selection function.  At present we generate a cubical volume that
encloses the survey volume and whose mean (constant) density matches
the maximum density required by the survey in question, as usually
done in mock catalogues constructed from $n$-body simulations (see
e.g. Cole et al 1998 for a detailed exposition).  This requires a lot
more galaxies than really needed.  The efficiency of {\tt PTHalos}
will be improved significantly by including these considerations from
the start, rather than imposing them at the end.  We hope to implement
this in the next version of the code, which will become publically
available.

\section{acknowledgments}

We thank Stephane Colombi, Lam Hui and Istvan Szapudi for useful
discussions.  RS thanks Fermilab, and RKS thanks NYU, for hospitality
at the initial and final stages of this project.  RKS is supported by
the DOE and NASA grant NAG 5-7092 at Fermilab.  The $n$-body
simulations used in Figs.~\ref{figSp}~and~\ref{figPkcova} were
produced using the Hydra $n$-body code (Couchman, Thomas, \& Pearce
1995). The $n$-body simulations used in Figs.~\ref{figpk}-\ref{figpkz}
are publically available at {\tt
http://www.mpa-garching.mpg.de/NumCos} and were carried out at the
Computer Center of the Max-Planck Society in Garching and at the EPCC
in Edinburgh, as part of the Virgo Consortium project.

\end{document}